\documentclass[aps,prd,showpacs,groupedaddress,floatfix,nofootinbib]{revtex4}
\usepackage{amsmath}
\usepackage{graphicx}
\usepackage{color}

\def\beq{\begin{eqnarray}}
\def\eeq{\end{eqnarray}}

\begin{document}

\title{The Schr\"odinger equation with piercings} 
\author{Paolo Amore}
\email{paolo.amore@gmail.com}
\affiliation{Facultad de Ciencias, CUICBAS,Universidad de Colima, \\
Bernal D\'{\i}az del Castillo 340, Colima, Colima, Mexico} 
\date{\today}

\begin{abstract}    
We show that the spectrum of the Schr\"odinger equation in two or higher dimensions
does not change when Dirichlet boundary conditions are enforced on a number of
isolated points inside the original domain (piercings). 
We have obtained the analytical solution for spherically symmetric state of the 
$d$-dimensional simple harmonic oscillator pierced at the origin. Results for the case
with multiple piercings are obtained numerically and  agree with the theoretical prediction. 
In the case of a two dimensional parabolic quantum dot with two electrons and a single piercing
in the origin we show that the energy of the ground state calculated to first order in perturbation 
theory goes over to the equivalent result in absence of piercing as the radius of the piercing becomes
 infinitesimal. Interestingly, we find that the leading finite size correction to the interaction energy 
 is negative while the corresponding correction to the single particle energies is (as expected) positive. 
For a finite radius of the piercing and above a critical coupling, the first term dominates the second 
and the total energy of the dot with piercing is lower. Unfortunately the critical coupling found for this 
example is nonperturbative. Finally we study configurations allowing bound states in the continuum, showing that 
bound states survive to the insertion of piercings and that their energy is unchanged when the radius 
of the piercing vanishes.
\end{abstract}
\pacs{03.30.+p, 03.65.-w}
\maketitle

\section{Introduction}
\label{intro}

It is a well known result that the frequencies of a membrane do not change when Dirichlet boundary conditions 
are enforced on a number of isolated points inside the domain of the membrane: this curious result, which at 
first seems to be at odds with common sense, was first conjectured by Lord Rayleigh long time ago~\cite{Ray}. 
More recently this problem has received attention in a number of papers, see for example 
\cite{Wang98,Gottlieb99,Laura99,Laura99b,Wang00,Wang01,Wang_Yu_01}. In particular, in ref.~\cite{Wang01} Wang has obtained 
a general expression for the correction to the frequencies of a membrane with $N$ internal circular regions of
infinitesimal radius where no vibration occurrs: these corrections are seen to decay proportionally to $1/\log r_0$, 
$r_0$ being the radius of the circular cores. In the present paper we show that these interesting results can be extended 
to Quantum Mechanics, at least for non--interacting particles. In Section \ref{sec1} we derive a general
formula for the correction to the energies of the Schr\"odinger equation in any number of dimensions, $d \geq 2$, 
with $N$ internal points of infinitesimal radius, where Dirichlet boundary conditions are enforced (we will refer to these 
points as to "piercings" throught all the paper). The case $d=2$ confirms the logarithmic behaviour of the corrections
already observed in the classical membrane vibration countepart, while a power-like behaviour is found for higher 
dimensions.  In Section \ref{sec2} we use the simple harmonic oscillator with a single piercing in the origin (pSHO) 
as the natural testing ground of the general results obtained: in this case we are able to confirm analytically 
the predictions of Section \ref{sec1}. The case with multiple piercings, on the other hand, is studied numerically, leading 
to a further confirmation of the results. An interesting question that may be asked is whether the same conclusions of Sections
\ref{sec1} and \ref{sec2} can be reached for $N$ interacting  quantum particles in $d$ dimensions: since the problem is clearly too difficult
to be attacked on general grounds, we have restricted our attention to a two dimensional parabolic quantum dot, with two 
electrons, which has been studied in ref.~\cite{Cifja04}. In this case we have been able to show that, to first order in perturbation
theory, the energy of the $2D$ quantum dot with a piercing in the origin is unchanged for $r_0 \rightarrow 0$; on the other
hand we have obtained that the leading order finite size correction to the energy is not definite positive and indeed may become
negative if the coupling of the Coulomb interaction is strong enough. Although the critical coupling where the change in sign
occurrs is too large to trust a first order perturbative calculation, we believe that this result opens the question on
whether the energy of the ground state of a quantum dot could be lowered adding one or more piercings of small but finite size.
In Section \ref{BIC} we have applied our results to the study of open systems supporting a 
bound state~\cite{Exner89,Schult89,Avishai91,Jaffe92,Trefethen06}, obtaining numerical results for the crossed wire configuration 
of Ref.\cite{Schult89} which confirm our predictions. In this way we prove that these bound states survive to the inclusion of piercings,
even for piercings of finite extension.
Finally, Section \ref{conclu} contains our conclusions.

\section{General considerations}
\label{sec1}

Consider the stationary Schr\"odinger equation (SSE) in $d$ spatial dimensions
\beq
- \frac{\hbar^2}{2m} \nabla^2_d \psi(x_1,\dots, x_d) + V(x_1,\dots,x_d) \psi(x_1,\dots,x_d) = E \psi(x_1,\dots,x_d)
\label{sch1}
\eeq
where $(x,y) \in {\cal D}$ and Dirichlet boundary conditions are 
enforced on $\partial {\cal D}$. 
Let $P_i = (x_1^{(i)}, \dots,x_d^{(i)})$ be a point internal to ${\cal D}$,  where Dirichlet b.c.
are also imposed. The index $i$ ranges from $1$ to $N$, i.e. the total number of  internal points with Dirichlet
bc. We call $\partial {\cal D}^* = \partial {\cal D} \ \cup \ \sum_{i=1}^N P_i$. 
In this case the SSE reads
\beq
- \frac{\hbar^2}{2m} \nabla^2_d \phi(x_1,\dots, x_d) + V(x_1,\dots, x_d) \phi(x_1,\dots, x_d) = E' \phi(x_1,\dots, x_d) \ ,
\label{sch2}
\eeq
$E'$ and $\phi$ being the new eigenvalue and wave function respectively. 
Assuming that $V(x_1,\dots, x_d)$ is real and using the two equations we can write
\beq
 (E-E') \phi^* \psi &=& - \frac{\hbar^2}{2m} \left[ \phi^* \nabla^2_d \psi- \psi \nabla^2_d \phi^*  \right]  = \vec{\nabla} \cdot \vec{J}  \ ,
\label{eq_1}
\eeq
where
\beq
\vec{J} \equiv  - \frac{\hbar^2}{2m} \left[ \phi^* \vec{\nabla}_d \psi- \psi \vec{\nabla}_d \phi^*\right] \ .
\eeq

After integrating eq.~(\ref{eq_1}) over $\cal D$ one has
\beq
E' = E - \frac{\int_{{\cal D}} \vec{\nabla} \cdot \vec{J} d^dx}{\int_{{\cal D}} \phi^*(x,y)\psi(x,y) d^dx} \ .
\label{eq_2}
\eeq

We use Gauss theorem to convert the integral over the $d$-dimensional volume of ${\cal D}$ in the numerator into 
an integral over $\partial {\cal D}^*$:
\beq
\int_{{\cal D}} \vec{\nabla} \cdot \vec{J} d^dx = \int_{\partial {\cal D}} \vec{J} \cdot d\vec{S}_{d-1} +
\sum_{i=1}^N \int_{P_i} \vec{J} \cdot d\vec{S}^{(i)}_{d-1}  \ ,
\eeq
where $S_{d-1}$ is the surface of ${\cal D}$ and $S_{d-1}^{(i)}$ is the surface of a $d$-sphere centered in $P_i$ with
infinitesimal radius. Since both $\psi$ and $\phi$ obey Dirichlet bc on $\partial {\cal D}$, the first integral vanishes.

On an $d-1$ sphere of infinitesimal radius $r_0$ around the point $P_i$ where Dirichlet boundary conditions are imposed one can express locally the
solution, $\phi$, as a linear combination of the regular solution $\psi$ and of a solution $\xi$ which diverges at $P_i$:
\beq
\phi \approx \psi + v_d \xi
\eeq
where $v_d$ is a constant to be determined by the condition $\phi(r_0) = 0$.

Keeping only the laplacian term in the Schroedinger equation, we approximate $\xi$ with the solution of the $d$-dimensional
Poisson equation, diverging at $P_i$:
\beq
\xi(r) = \left\{\begin{array}{cc}
\log r &  d=2 \\
\frac{1}{r^{d-2}} & d>2 
\end{array} \right.
\eeq

We therefore obtain
\beq
v_d(r_0) = \left\{\begin{array}{cc}
-\frac{\psi(r_0)}{\log r_0} &  d=2 \\
-\psi(r_0) r_0^{d-2} & d>2 
\end{array} \right.
\eeq

After taking into account these observations, and neglecting the terms which vanish after integration we have
\beq
\int_{{\cal D}} \vec{\nabla} \cdot \vec{J} d^dx &=&  \sum_{i=1}^N \frac{\hbar^2}{2m} v_d^{(i)} \psi(r_0^{(i)}) \int_{P_i} 
\vec{\nabla}\xi^* \cdot d\vec{S}^{(i)}_{d-1} = 
- \sum_{i=1}^N \frac{\hbar^2}{2m} v_d^{(i)} \psi(r_0^{(i)}) \vec{\nabla}\xi^*\cdot\hat{r}|_{r_0^{(i)}}  S^{(i)}_{d-1} \ ,
\eeq
where $S^{(i)}_{d-1}$ is the surface of a $d-1$ sphere, which is given by
\beq
S^{(i)}_{d-1} = \frac{2 \pi^{d/2} r_0^{d-1}}{\Gamma(d/2)} \ .
\eeq

Therefore we may write:
\beq
\int_{{\cal D}} \vec{\nabla} \cdot \vec{J} d^dx &=&  \left\{\begin{array}{cc}
 \sum_{i=1}^N \frac{\hbar^2 \pi}{m}  \frac{\psi^2(r_0^{(i)})}{\log r_0^{(i)}}  &  d=2 \\
\sum_{i=1}^N \frac{\hbar^2}{m}  \psi^2(r_0^{(i)}) (d-2) (r_0^{(i)})^{d-2} \frac{\pi^{d/2}}{\Gamma(d/2)}   &  d>2 
\end{array}\right.
\label{eq_3}
\eeq

Let us now come to the denominator in eq.~(\ref{eq_2}): since we expect that the wave function in the presence of piercing
be sensibly modified only in the close neighborhood of the points themselves, and the wave functions being normalized, 
we may argue that $\int_{{\cal D}} \phi^*(x,y)\psi(x,y) d^dx \approx 1$. The leading behaviour of the energy for $r_0\rightarrow 0$
can thus be obtained from eq.~(\ref{eq_3}).

An alternative --qualitative -- argument to convince the reader of the isospectrality could be the following: 
consider the wave function
\beq
\phi(x_1, \dots, x_d) = \psi(x_1,\dots,x_d) \ \Xi(x_1,\dots, x_d) 
\label{asym}
\eeq
where 
\beq
\Xi(x_1,\dots, x_d) \equiv \prod_{i=1}^N \xi^{(i)}(x_1,\dots,x_d)
\eeq
and $\xi^{(i)}$ is the solution to the homogeneous Poisson equation singular at a point $P_i$ internal
to ${\cal D}$ and vanishing on an infinitesimal circle of radius $r_0$. $\psi(x_1,\dots,x_d)$ is the solution in absence of piercing.
We substitute $\phi$ in eq.~(\ref{sch2}) and we obtain the equation:
\beq
-\frac{\hbar^2}{2m} \ \left[2 \vec{\nabla} \psi \cdot \vec{\nabla} \Xi + \psi \nabla^2 \Xi \right]+
(E-E') \phi(x_1,\dots,x_d)  = 0 \ .
\eeq

If we multiply this equation by $\phi$ and integrate over the $d$ dimensional volume we have
\beq
(E-E') = \frac{\hbar^2}{2m} \ \int_{{\cal D}^*} \phi  \left[2 \vec{\nabla} \psi \cdot \vec{\nabla} \Xi + \psi \nabla^2 \Xi \right] d^dx \ .
\eeq

Since the expression in the square brackets is a function which is singular at all the internal points $P_i$, 
we expect that the main contribution to the integral comes from these regions. Around these points, however,
$\phi=0$, and therefore we may expect that 
\beq 
\Delta E = \int_{{\cal D}^*} \phi  \left[2 \vec{\nabla} \psi \cdot \vec{\nabla} \Xi + \psi \nabla^2 \Xi \right] d^dx \approx 0 \ ,
\eeq
implying that $E' \approx E$. 

We can also provide a more direct proof of this statement by considering the special case of a single piercing in two dimensions, 
which can then be generalized to $N$ piercings in $d$ dimensions. In this case  eq.~(\ref{asym}) reads
\beq
\phi(x,y) = \psi(x,y) \ \log \frac{r}{r_0} \ .
\eeq

As it stands this expression is not normalized and we should rather write
\beq
\tilde{\phi}(x,y) = \frac{\psi(x,y) \ \log \frac{r}{r_0}}{\sqrt{\int_{{\cal D}^*} |\psi(x,y)|^2 \log^2 \frac{r}{r_0} dxdy}} \ .
\eeq

If we now take the limit $r_0 \rightarrow \infty$,for $r > r_0$, we have
\beq
\lim_{r\rightarrow r_0} \tilde{\phi}(x,y) = \psi(x,y)
\eeq
which supports the conclusions previously reached.

Since the results obtained in this Section concern the inclusion of isolated Dirichlet points
in the $d$ dimensional domain where the SSE is defined, the reader may wonder at this point if  it is possible 
to find regions of dimension $0<d'<d-2$ which leave the spectrum invariant when Dirichlet bc are 
enforced. In this case the application of Gauss theorem must be specific to the problem, since $S_{d-1}^{(i)}$ 
will be now the infinitesimal surface enclosing the Dirichlet region. We may however convince ourselves that
it is indeed possible by looking at a specific example: consider the Hamiltonian for a $d\geq 3$ problem, which is separable 
in the form $\hat{H} = \hat{H}_{1\dots k} + \hat{H}_{k+1,\dots,d}$, where $k \geq 2$. 
Here $\hat{H}_{1\dots k}$ contains the coordinates $x_1,\dots, x_k$ and their 
conjugate momenta, while $\hat{H}_{k+1,\dots,d}$ contains the remaining coordinates and momenta. In this case the solution of the
problem is obtained as the direct product of the solutions of $\hat{H}_{1\dots k}$ and $\hat{H}_{k+1,\dots,d}$ respectively. Therefore, 
adding $N$ piercings of infinitesimal size to $\hat{H}_{1\dots k}$ can be interpreted as adding $N$ regions of dimension $d-k$ with Dirichlet 
boundary conditions to the original domain. Since the spectrum of $\hat{H}_{1\dots k}$ is unchanged under this operation, we conclude
that the same is true for the total hamiltonian $\hat{H}$.

\section{The simple harmonic oscillator in $d$ dimensions pierced in the origin}
\label{sec2}

The simple harmonic oscillator in $d>2$ dimensions is the ideal test of the results presented 
in the previous section, given that analytical results can be obtained. The SSE in this case reads
\beq
\left[- \frac{\hbar^2}{2\mu} \nabla_d^2  + \frac{1}{2} \mu \omega^2 (x_1^2+
\dots + x_d^2) - E\right] \psi(x_1,\dots, x_d) = 0 \ , \nonumber 
 \eeq
where $\nabla_d^2$ is the Laplacian operator in $d$ dimensions. For simplicity
we study only spherically symmetric states with zero angular momentum and therefore 
consider only the radial part of the Laplacian ($\nabla_d^2 \rightarrow \frac{\partial^2}{\partial r^2} + 
\frac{d-1}{r} \frac{\partial}{\partial r}$).~\footnote{Notice as well that the 
remaining states authomatically comply with the Dirichlet boundary condition 
at the origin and therefore both their energies and wave functions are unaffected.}

After introducing the standard dimensionless variables 
$\rho \equiv \sqrt{\mu \omega/\hbar} \ r$ and $\epsilon \equiv E/\hbar\omega$~\footnote{Actually, since the numerical
results discussed in this paper are obtained for unit mass and setting $\hbar = \omega = 1$, the use of $\rho$ or $r$ 
is equivalent.} we write $\psi(\rho) = e^{-\rho^2/2} \xi(\rho)$ and obtain the differential 
equation
\beq
\xi''(\rho) +  \left( \frac{d-1}{\rho} - 2 \rho \right) \xi'(\rho)+
\left(2\epsilon - d \right) \xi(\rho) = 0 \ .
\eeq

\begin{figure}
~\bigskip\bigskip
\begin{center}
\includegraphics[width=8cm]{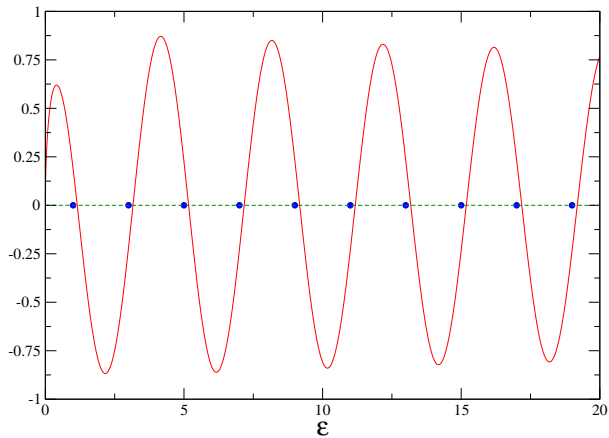}
\caption{$\Omega(\epsilon,\rho_0) \equiv \frac{U\left(\frac{2-2 \epsilon}{4},1,\rho_0^2\right)}{\log_{10} 1/\rho_0 \ \sqrt{\epsilon} \ \Gamma \left( \frac{\epsilon}{2}\right)}$
for $\rho_0 = 10^{-3}$ as a function of $\epsilon$. (color online)}
\bigskip
\label{fig1}
\end{center}
\end{figure}

We wish to solve this equation subject to the conditions $\xi(\rho_0)=0$
and $\lim_{\rho\rightarrow\infty} \psi(\rho)=0$, so that the wave function be square
integrable on the domain. After enforcing the first condition we obtain the solution:
\beq
\xi(\rho) &=& \left[ U\left(\frac{d-2 \epsilon}{4},\frac{d}{2},\rho_0^2\right) 
L_{\frac{\epsilon}{2}-\frac{d}{4}}^{\frac{d}{2}-1}\left(\rho ^2\right) 
-  U\left(\frac{d-2 \epsilon }{4},\frac{d}{2},\rho
   ^2\right) L_{\frac{\epsilon }{2}-\frac{d}{4}}^{\frac{d}{2}-1}\left(\rho_0^2\right)\right] \nonumber \\
\label{exact0}
\eeq
where $U$ and $L$ are the hypergeometric $U$ and the Laguerre $L$ functions.
Since the Laguerre function blows up at infinity the full solution also blows up unless the condition
\beq
U\left(\frac{d-2 \epsilon}{4},\frac{d}{2},\rho_0^2\right) = 0 
\label{exact}
\eeq
is met. As we see in Fig.~\ref{fig1}, which displays $\Omega(\epsilon,\rho_0) \equiv \frac{U\left(\frac{2-2 \epsilon}{4},1,\rho_0^2\right)}{\log_{10} 1/\rho_0 \ 
\sqrt{\epsilon} \ \Gamma \left( \frac{\epsilon}{2}\right)}$ for $\rho_0 = 10^{-3}$, the zeroes of this function are slightly displaced to the right with respect 
to the unperturbed values, represented with circles. Notice that for specific values of the energy, the expression for the wave function simplify: for example, 
in two dimensions, for $\epsilon = 2 k+1$, with $k$ integer, one goes over to the wave functions of the excited states of the unpierced oscillator 
(this corresponds to setting $\rho_0$ in the largest positive node of such wave function). For $\epsilon = 3$ one has
$\xi(\rho) \propto (1-\rho^2)$, corresponding to $\rho_0 = 1$; for $\epsilon = 5$, one has $\xi(\rho) \propto (2-4 \rho^2+\rho^4)$,
corresponding to $\rho_0 = \sqrt{2+\sqrt{2}}$.

In particular we are interested in knowing the behaviour of the energies of the ground state for $\rho_0 \rightarrow 0$: 
for $d=2$ we obtain that $\epsilon^{(2)} \approx 1 - 1/\log \rho_0$, whereas for $d>2$ we have
$\epsilon^{(d)} \approx d/2 + c^{(d)} \ \rho_0^{d-2}$, where $c^{(d)}= 2/\Gamma(d/2-1)$. 
Notice that this expression for $c^{(d)}$ coincides with the general expression in eq.~(\ref{eq_3}).

We can now use eq.~(\ref{exact0}) to extract the dominant behaviour of the exact solution as $\rho_0 \rightarrow 0$.
In the limit $r \rightarrow 0$, for the ground state,  the confluent hypergeometric function behaves as
\beq
U(a,\frac{d}{2},r^2) \approx  \left\{
\begin{array}{cc}
- 2 a \log r & d=2\\
a \frac{\Gamma(d/2-1)}{r^{d-2}} & d>2
\end{array}
\right.
\eeq
which  confirms eq.~(\ref{asym}). 

\begin{figure}
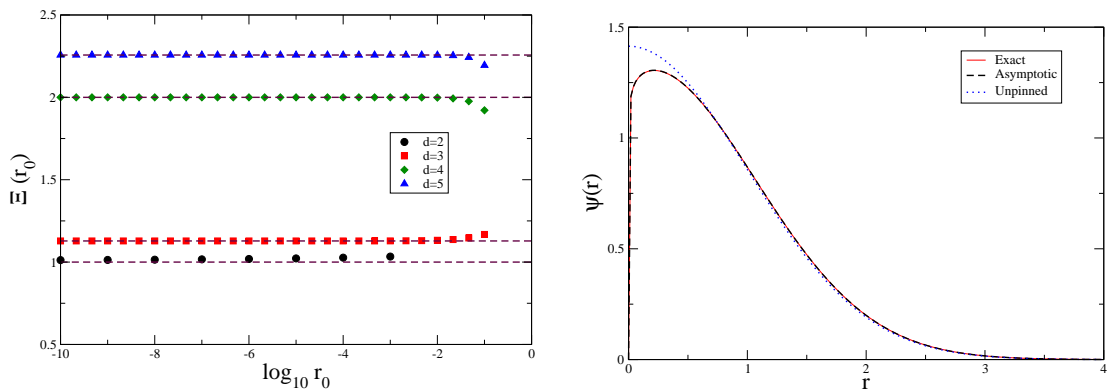

~\bigskip\bigskip
\begin{center}
\includegraphics[width=7cm]{fig2a.eps} 
\hskip .5cm
\includegraphics[width=7cm]{fig2b.eps}
\caption{Left plot:$\Xi(r_0) \equiv (\epsilon_d -d/2)/r_0^{d-2}$ for the ground state of the simple harmonic oscillator in $d$ dimensions, pierced at the origin.
Right plot:Wave function for the ground state of the sho: the solid and dashed lines correspond to the exact and asymptotic solutions
for the sho pierced at the origin; the dotted line is the unpierced sho. The parameter $r_0 = 10^{-10}$ is used in the first two curves.  (color online)}
\bigskip
\label{fig2}
\end{center}
\end{figure}

In the left plot of Fig.\ref{fig2} we display the quantity $\Xi(r_0) \equiv (\epsilon_d -d/2)/r_0^{d-2}$ for the ground state
of the SHO pierced in the origin in $d=2,3,4,5$ dimensions for small values of  $r_0$. The horizontal lines are the 
coefficients $c_d$ obtained directly from the exact solution and from the general formula, given in the previous section.
In the right plot of Fig.\ref{fig2} we compare the exact solution of eq.~(\ref{exact0}) and the asymptotic one of eq.~(\ref{asym}), 
both corresponding to a piercing radius $r_0 = 10^{-10}$, with the solution for the oscillator without piercing.

\begin{figure}
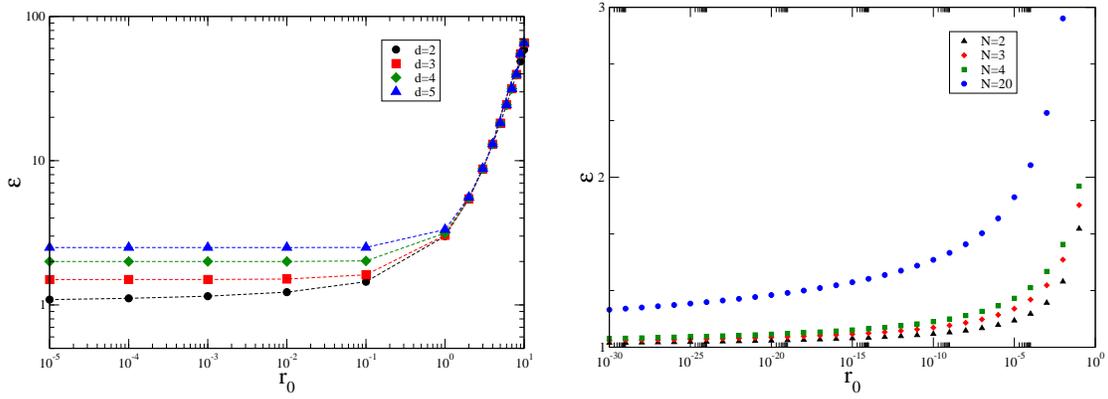

~\bigskip\bigskip
\begin{center}
\includegraphics[width=7cm]{fig3a.eps}
\hskip 0.5cm
\includegraphics[width=7cm]{fig3b.eps}
\caption{Left plot: Energy of the ground state of the $d$-dimensional simple harmonic oscillator as a function of the radius $r_0$ of the 
internal region where the wave function vanishes. The symbols correspond to the numerical solution of the exact equation, whereas
the dashed lines correspond to the variational estimate. 
Right plot:Energy of the ground state of the two dimensional quantum dot with $N$ piercings uniformly distributed at a distance $d=1/3$ from the origin,
calculated using the asymptotic wave function. We use $\hbar = m = \omega = 1$ and plot the energy as a function of the radius $r_0$.  (color online)}
\bigskip
\label{fig3}
\end{center}
\end{figure}

The energy of the ground state of the pierced sho can be estimated precisely even in the case where 
the radius of the pierced region is sizeable using a variational approach. In this case
we use a trial wave function
\beq
\psi(\rho) = N \ e^{-\rho^2/2} \ \log \left(\frac{\rho}{\rho_0} \right) \ 
\left(1+a \rho^b\right)
\eeq
for $d=2$ and
\beq
\psi(\rho) = N \ e^{-\rho^2/2} \ \left(\frac{1}{\rho^{d-2}} - \frac{1}{\rho_0^{d-2}}\right) \ 
\left(1+a \rho^b\right)
\eeq
for $d>2$. Here $N$ is a normalization constant and $a$ and $b$ are variational parameters.
In the left part of Fig.~\ref{fig3} we show the energy of the ground state of the pierced sho, as a function of the pinning 
radius, $r_0$, for $d=2,3,4,5$. In the limit $r_0 \rightarrow 0$ the energies tend to the energy of the 
unpierced sho, whereas for $r_0 > 1$ the curves approach a universal behaviour, which is dominated by the
potential energy. Notice that the symbols correspond to the energy calculated numerically 
using the exact expression,  while the lines correspond to the variational results. 

We briefly address the problem of a SHO with $N$ piercings in two dimensions. Although this problem cannot be solved exactly 
for an arbitrary radius of the piercing, its spectrum should still go over to the unperturbed spectrum as the radius of the piercing
goes to zero, with a logarithmic strength which is given by eq.~(\ref{eq_3}). In Table \ref{tab1} we compare the
numerical results for the energy and the leading coefficient $c^{(2)}$, obtained by fitting the energy obtained
using the expectation value of the hamiltonian in the asymptotic wave function for $r_0$ going from $10^{-30}$ to $10^{-20}$,
with the theoretical predictions obtained in Section \ref{sec1}. The overlap between the asymptotic and unperturbed wave functions
has been calculated numerically for $r_0=10^{-50}$: this overlap, as anticipated, is fairly close to $1$ and simply assuming it
$1$ would lead to $c^{(2)} = N \ e^{-1/9} \approx 0.894839 N$.

\begin{figure}
~\bigskip\bigskip
\begin{center}
\includegraphics[width=9cm]{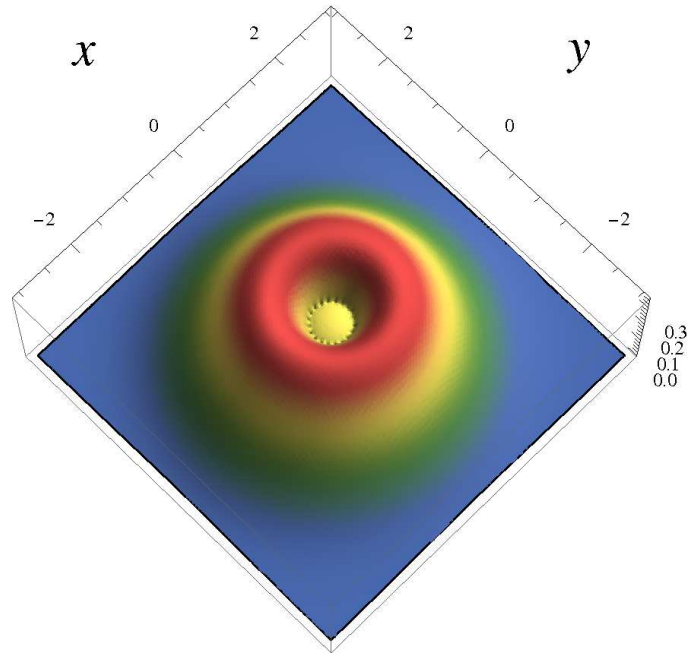}
\caption{Asymptotic wave function for the SHO pierced on $20$ points uniformly distributed on a circle of radius $R=1/3$. In this case $r_0 = 10^{-10}$.
 (color online)}
\bigskip
\label{fig4}
\end{center}
\end{figure}

The right plot of Fig.~\ref{fig3} displays the energy of the ground state of the SHO with $N$ piercings uniformly distributed 
on a circle of radius $R=1/3$. The plot of the asymptotic wave function for the SHO with $20$ piercing of radius $r_0 = 10^{-10}$
is shown in Fig.\ref{fig4}.

\begin{table}
\caption{\label{tab1} Theoretical and numerical results for the energy and the leading coefficient $c^{(2)}$ of a SHO which contains 
$N$ piercings uniformly distributed  on a circle of radius $1/3$.}
\begin{ruledtabular}
\begin{tabular}{ccccc}
$N$ & $\epsilon_{th}$ & $\epsilon_{FIT}$ &  $c_{th}^{(2)}/N$  & $c_{FIT}^{(2)}/N$  \\ 
\hline
2  &  1 &  0.9999977 & 0.8948766 &  0.8950938  \\
3  &  1 &  0.9999314 & 0.8948683 &  0.8987901 \\
4  &  1 &  0.9999960 & 0.8949873 &  0.8950401  \\
10 &  1 &  0.9998796 & 0.8956917 &  0.8976956  \\
20 &  1 &  0.9993188 & 0.8983882 &  0.9036415 \\
\end{tabular}
\end{ruledtabular}
\bigskip\bigskip
\end{table}

\section{Ground state of a two dimensional parabolic quantum dot}
\label{sec3}

We now consider  a two electron system confined by a 2D parabolic potential in a zero magnetic field. Ref.~\cite{Cifja04} contains an analytical expression 
for the first order contribution to the ground state energy of this system in nondegenerate perturbation theory. We will first review the main steps of this 
calculation for the unpierced quantum dot and then extend the calculation to the quantum dot pierced in the origin.

The hamiltonian for this problem is
\beq
\hat{H}(\vec{r_1},\vec{r_2}) = \frac{\hat{p}_1^2}{2\mu} + \frac{\mu \omega^2 r_1^2}{2} + \frac{\hat{p}_2^2}{2\mu} + \frac{\mu \omega^2 r_2^2}{2} +
\frac{e^2}{|\vec{r}_1-\vec{r}_2|} \ ,
\eeq
where the last term provides the Coulomb repulsion between the electrons. 

In the absence of this term the total wave function is obtained as the direct product of the single particle wave functions of a simple harmonic oscillator,
which are given by:
\beq
\Phi_{n,m}(r,\phi) = N_{nm}  \frac{e^{i m\phi}}{\sqrt{2\pi}} \ (\alpha r)^{m} \ e^{-\alpha^2 r^2/2} L_n^{|m|}(\alpha^2 r^2) , 
\eeq
where $L_n^m$ are the associated Laguerre polynomials. $N_{nm} = \sqrt{2n! \alpha^2/(n+|m|)!}$ is the normalization factor and $\alpha  \equiv \sqrt{\mu\omega/\hbar}$.
The single particle energies are given by
\beq
E_{nm} = \hbar \omega (2 n+|m|+1),
\eeq
with $n=0,1,\dots$ and $m = 0,\pm 1,\pm 2,\dots$. The single particle wave function for the ground state is simply
\beq
\Phi_{00}(r,\phi) = \frac{\alpha}{\sqrt{\pi}} e^{-\alpha^2 r^2/2} \ .
\eeq

In the ground state of the two particle system the electrons form a spin singlet and therefore the orbital wave function is symmetric in the electron coordinates:
\beq
\Psi(\vec{r}_1,\vec{r}_2)  = \Phi_{00}(\vec{r}_1) \Phi_{00}(\vec{r}_2) \ \left[\frac{\chi_{+,1} \chi_{-,2}-\chi_{+,2} \chi_{-,1}}{\sqrt{2}} \right] \ .
\eeq

Treating the Coulomb repulsion between the electrons as a perturbation, the authors of Ref.~\cite{Cifja04} have obtained an analytic expression for 
the first order correction in nondegenerate perturbation theory:
\beq
\Delta E^{(1)} &=& \left( \frac{\alpha^2}{\pi}\right)^2 \ \int d^2r_1 \int d^2 r_2  e^{-\alpha^2 (r_1^2+r_2^2)} \ \frac{e^2}{|\vec{r}_1-\vec{r}_2|}  
=  e^2 \alpha \sqrt{\frac{\pi}{2}} \ ,
\eeq
using the identity~\footnote{Because the ground state is spherically symmetric only the $m=0$ term in this expression contributes.}
\beq
\frac{1}{|\vec{r}_1-\vec{r}_2|} = \sum_{m=-\infty}^{+\infty} \int_0^\infty dk e^{i m (\phi_1-\phi_2)} \ J_m(k r_1) \ J_m(k r_2)  \ .
\eeq

We may now discuss the same problem in the presence of a piercing in the origin. As previously found, the single particle wave function for the
ground state in presence of piercing of infinitesimal radius $r_0$ is simply given by
\beq
\tilde{\Phi}_{00}(\vec{r}) = \tilde{N}_{00} \ e^{-\alpha^2 r^2/2} \ \log \frac{r}{r_0}  . 
\eeq
The normalization constant is expressed in term of the Meijer $G$ function as
\beq
\tilde{N}_{00} = \alpha \sqrt{\frac{2}{\pi \ G_{2,3}^{3,0}\left(\alpha^2 {r_0}^2 \left|
\begin{array}{c}
 1,1 \\
 0,0,0
\end{array}
\right.\right)}} \ ,
\eeq
and $\tilde{N}_{00} \rightarrow - \frac{\alpha}{\sqrt{\pi} \log \alpha r_0} +  \frac{\alpha \gamma}{2\sqrt{\pi} \log^2 \alpha r_0}$ for $r_0 \rightarrow 0$.

The perturbative correction to the ground state energy is therefore given by
\beq
\Delta \tilde{E}^{(1)} &=& e^2 \tilde{N}_{00}^4  \int d^2r_1 \int d^2 r_2 \ \log^2 \frac{r_1}{r_0} \log^2 \frac{r_2}{r_0} \ e^{-\alpha^2 (r_1^2+r_2^2)} \ 
\frac{e^2}{|\vec{r}_1-\vec{r}_2|}  \nonumber \\
&=& e^2 \tilde{N}_{00}^4  \int_0^\infty dk \left[\int d^2r  \log^2 \frac{r}{r_0}  e^{-\alpha^2 r^2} \ J_0(kr) \right]^2 
\label{eqdot}
\eeq

\begin{figure}
~\bigskip\bigskip
\begin{center}
\includegraphics[width=8cm]{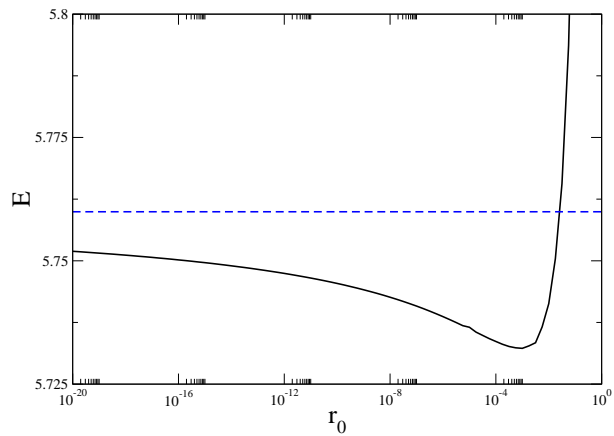}
\caption{Energy of the ground state of the two dimensional quantum dot with pinning to first order in perturbation theory for $\lambda = 3$ and 
$\hbar = m = \omega = 1$, as a function of the radius $r_0$. The horizontal line is the equivalent result in the absence of pinning.  (color online)}
\bigskip
\label{fig5}
\end{center}
\end{figure}

In order to evaluate the integral in the square bracket we use the series representation
\beq
J_0(z) =  \sum_{j=0}^\infty \frac{(-1)^j}{j!^2} \ \left( \frac{z}{2} \right)^{2j} \ , 
\label{bessel}
\eeq
and evaluate the integral
\beq
{\cal I}_{j} &\equiv& \int_{r_0}^\infty dr \ r^{2j+1} \log^2 \frac{r}{r_0}  e^{-\alpha^2 r^2}   =
\frac{1}{4} \alpha^{-2 (j+1)} G_{3,4}^{4,0}\left(\alpha^2 {r_0}^2\left|
\begin{array}{c}
 1,1,1 \\
 0,0,0,j+1
\end{array}
\right.\right) \nonumber \\
&\approx& \frac{j! }{2  \alpha^{2 (j+1)}} \ 
 \left[ \log ^2(\alpha {r_0})-\log (\alpha {r_0}) \psi^{(0)}(j+1)+ \dots \right]
\eeq
where $\psi^{(0)}(j+1)$ is the polygamma function given by $\psi^{(0)}(z) = \sum_{k=1}^\infty \left[ \frac{1}{k} - \frac{1}{k+z-1}\right] - \gamma$.
We use this expression, together with eq.~(\ref{bessel}), to write eq.~(\ref{eqdot}) as
\beq
\Delta \tilde{E}^{(1)} &\approx& 4\pi^2 e^2 \tilde{N}_{00}^4 \int_0^\infty dk \frac{1}{4\alpha^4} e^{-\frac{k^2}{2 \alpha^2}} 
\left[ \log^4 \alpha r_0 -2 \left(\Gamma \left(0,-\frac{k^2}{4 \alpha^2}\right)\right.\right. \nonumber \\
&+& \left.\left.\log \left(-\frac{k^2}{4 \alpha^2}\right)\right)^2 \log ^3(\alpha {r_0})
+ \dots \right] 
= e^2 \alpha \left[ \sqrt{\frac{\pi}{2}} + \frac{\kappa-\sqrt{2\pi} \gamma}{\log \alpha r_0} +\dots \right]
\eeq
where $\kappa \equiv - 2  \int_0^\infty dk e^{-\frac{k^2}{2}} \left(\Gamma \left(0,-\frac{k^2}{4 }\right)+\log \left(-\frac{k^2}{4}\right)\right)
\approx  2.240700135$. This equation tells us that in the limit $r_0 \rightarrow 0$ the interaction energy 
to first order in PT is exactly the same as in the case of the quantum dot without piercing, while the leading finite size correction to this 
energy is negative. In other words, at least to first order in PT, the presence of a piercing lowers the interaction energy.
Combining this expression with the leading order expression for the single particle energy we obtain the total energy 
to first order in PT goes as
\beq
E_{TOT} \approx   \hbar \omega \left\{ 2 +  \lambda \sqrt{\frac{\pi}{2}} - \frac{1}{\log \alpha r_0} 
\left[ 2 - (\kappa-\sqrt{2\pi} \gamma) \lambda  \right] \right\}
\eeq
where $\lambda \equiv \frac{e^2\alpha}{\hbar \omega}$, using the notation of  Ref.~\cite{Cifja04}.
For $\lambda > \lambda_{min} = \frac{2}{\kappa-\sqrt{2\pi} \gamma} \approx 2.51941$ the total energy of the system is lower in presence of a piercing,  
for sufficiently small values of $r_0$. On the other hand we know that for large $r_0$, the single particle energy grows rapidly, thus suggesting
the presence of a minimal energy at a finite $r_0$.
Fig.~\ref{fig5} confirms this expectation and shows the energy of the ground state of the two dimensional quantum dot with piercing for $\lambda = 3$ 
calculated to first  order in perturbation theory, using the single particle wave functions of eq.~(\ref{exact0}). 
The interaction energy in this case is evaluated numerically. 

Notice that a fit of the numerical results obtained for $10^{-20} < r_0 < 10^{-10}$
yields
\beq
E^{(1)}_{FIT}(r_0)  \approx \alpha e^2 \ \left[ 1.25322 + \frac{0.79415}{\log \alpha r_0}+\frac{0.210805}{\log^2 \alpha r_0}-\frac{0.572547}{\log^3 \alpha r_0} 
+ \dots \right]
\eeq
which agrees remarkably well with the theoretical result previously obtained
\beq
E^{(1)}_{th}(r_0) = e^2 \alpha \left[\sqrt{\frac{\pi}{2}} + \frac{\kappa-\sqrt{2\pi} \gamma}{\log \alpha r_0} + \dots  \right] \approx 
e^2 \alpha \left[ 1.25331 + \frac{0.793835}{\log \alpha r_0}+ \dots \right] \ .
\eeq

\section{Bound states in the continuum}
\label{BIC}

We wish now to apply the general results obtained in Section \ref{sec1} to the case of the Helmholtz equation on 
domains which extend to infinity, but consist of wires with crossings and/or bendings. It is  well known that 
in this case  the spectrum of the Laplacian contains one or more bound states, depending on the number of crossings and bendings, 
below the threshold of the continuum. 
For example, Exner and Seba first showed in  \cite{Exner89} that a smoothly curved waveguide can support a bound state; 
Schult et al. have reached a similar conclusion in \cite{Schult89} for a different configuration consisting of two crossed 
wires, of infinite length. Avishai and collaborators have also proved the existence of a bound state in the broken strip 
configuration for arbitrarily small angles (see \cite{Avishai91}), while Goldstone and Jaffe \cite{Jaffe92} have given a 
variational proof of the existence of a bound state for an infinite tube with bendings in two and three dimensions. 

The general results obtained in our Section \ref{sec1} suggest that when a number of piercings of infinitesimal radius  is added to a crossed wire 
or to a bent waveguide, the energy of the bound state remains precisely the same although the corresponding wave function changes. 
If the radius $r_0$ of the piercing is now made finite, still one expects the bound state to survive, up to some critical value of $r_0$.
Notice that the considerations made at the end of Section \ref{sec1} also indicate the possibility to extend the same results to filaments inside 
three dimensional bent tubes.

In this Section we provide an explicit confirmation of our general results, by studying  the crossed wire configuration of Ref.\cite{Schult89}.
In analogy with Ref.~\cite{Schult89} we have decided to study this problem using two methods: the first method is based on a collocation
approach which allows one to obtain a numerical solution to the Helmholtz equation;  the second method uses an expansion in a complete set of 
solutions of the Helmholtz equation on the five domains which compose the global domain (shown in Fig.1 of Ref.\cite{Schult89}). The explicit
expressions for these sets of functions may be found in the paper of Schult et al.

Let us first briefly describe the first method. In this case we have discretized the Helmholtz equation on a uniform grid
using a set of functions, called Little Sinc Functions (LSF), derived in Ref.~\cite{Amore07}. These functions have recently been
applied in \cite{Amore08} to the study of vibration of membranes of arbitrary shapes. Although we refer the reader to Ref.~\cite{Amore08}
for the technical details concerning the implementation of this method, we just mention that the inclusion of piercings inside the domain
is handled with extreme simplicity in this approach, provided that the piercing falls on one of the point forming the mesh. As a matter of fact, 
each LSF is an approximate representation of a Dirac delta function, peaked on one of the points of the mesh: the exclusion of a point from the 
mesh is therefore obtained by eliminating the corresponding LSF from the set of functions used to discretize the Hamiltonian.

\begin{figure}
~\bigskip\bigskip
\begin{center}
\includegraphics[width=8cm]{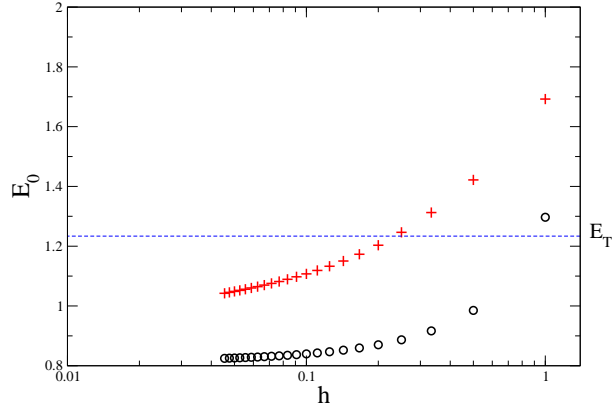}
\caption{Energy of the ground state of the crossed wired of width $w=1$ calculated using the collocation method as a function of the
grid spacing $h$. The circles are the results for the crossed wire without piercing and the crosses are the results for the crossed wire pierced in the 
origin. The horizontal line is the threshold for the continuum. (color online)}
\bigskip
\label{fig6}
\end{center}
\end{figure}

Using the collocation method we have solved numerically the Helmholtz equation on the crossed wire domain, assuming that the arms have a width $w=2$  
(we have also set $\hbar = m = 1$ in our calculation). Although the arms have infinite extension, we have used arms of finite length $L=10$, in other words we 
have limited the crossed wire to the interior of a square of sides $20 \times 20$. This choice is expected to affect very mildly the lowest energy state 
which is localized at the crossing. The energy obtained in this way provides in any case an upper bound to the true energy.

In Fig.~\ref{fig6} we display the energy of the ground state obtained numerically with the LSF, for the case of a crossed wire without
piercings (circles) and for a crossed wire with a piercing in the origin. The horizontal line is the threshold energy for the continuum.
These results are obtained for specific values of the grid spacing, $h$, for which the boundary of the crossed wire falls precisely on the 
mesh: as discussed in Ref.~\cite{Amore08}, in this case one obtains a monotonous sequence of values which converge to the exact result from 
above. We have extracted the continuum limit fitting the two sets of data with 
\beq
E_0^{(a)} &=& a_0 \ \frac{1+ a_1 h^{a_2}}{1+ a_3 h^{a_4}} 
\label{lsq1}
\eeq
for the case without piercing and
\beq
E_0^{(b)} &=& b_0 \ \frac{1+ b_1/\log h + b_2 h^{b_3}}{1+ b_4/\log h + b_5 h^{b_6}} \ ,
\label{lsq2}
\eeq
for the case with piercing. The coefficients $a_i$ and $b_i$ are obtained by a least square procedure.
The last row of the Table contains the results obtained from the coefficients $a_0$ and $b_0$ of the fits above using a least square 
procedure: the results for the two cases are almost degenerate and agree with the result quoted in Ref.~\cite{Schult89}, i.e.
$k w = \frac{w \sqrt{2 m E}}{\hbar}\approx 0.812 \pi$.

\begin{table}
\caption{\label{tab2} Ground state energy of the crossed wire obtained with the collocation method.
$E_0$ ($E_0^{pierced}$) is the energy of the configuration without (with) piercing. The last row are
the energies obtained using eqs.~(\ref{lsq1}) and (\ref{lsq2}) with a least square approach.}
\begin{ruledtabular}
\begin{tabular}{cccccc}
$h$ & $E_0$ & $E_0^{pierced}$ & $h$ & $E_0$ & $E_0^{pierced}$ \\
\hline
$1$            & 1.296833 & 1.692234 & $\frac{1}{12}$ & 0.834904 & 1.089169\\
$\frac{1}{2}$  & 0.985217 & 1.421753 & $\frac{1}{13}$ & 0.833134 & 1.081897\\
$\frac{1}{3}$  & 0.916621 & 1.312652 & $\frac{1}{14}$ & 0.831632 & 1.075505\\
$\frac{1}{4}$  & 0.886707 & 1.246377 & $\frac{1}{15}$ & 0.830342 & 1.069828\\ 
$\frac{1}{5}$  & 0.870046 & 1.203231 & $\frac{1}{16}$ & 0.829222 & 1.064742\\
$\frac{1}{6}$  & 0.859464 & 1.172968 & $\frac{1}{17}$ & 0.828241 & 1.060153\\
$\frac{1}{7}$  & 0.852166 & 1.150460 & $\frac{1}{18}$ & 0.827374 & 1.055983\\
$\frac{1}{8}$  & 0.846836 & 1.132966 & $\frac{1}{19}$ & 0.826604 & 1.052173\\
$\frac{1}{9}$  & 0.842779 & 1.118906 & $\frac{1}{20}$ & 0.825914 & 1.048673\\
$\frac{1}{10}$ & 0.839590 & 1.107308 & $\frac{1}{21}$ & 0.825294 & 1.045443\\
$\frac{1}{11}$ & 0.837019 & 1.097539 & $\frac{1}{22}$ & 0.824732 & 1.042450\\
\hline
LSQ            & 0.813917 & 0.813737 &  & & \\
\end{tabular}
\end{ruledtabular}
\bigskip\bigskip
\end{table}

Let us now describe the second method. As done in  \cite{Schult89} we express the solution in each of the five regions
as a linear combination of elementary solutions fulfilling the Helmholtz equation:
\beq
\psi_I(x,y) &=& \sum_{n=0}^{N-1}  b_n \cosh \frac{\rho_n w}{2}  e^{\rho_n (w/2-x)} \ \cos \frac{(2n+1) \pi y}{w} \\
\psi_V(x,y) &=& \sum_{n=0}^{N-1}  b_n \left[ \cosh \rho_n x  \ \cos  \frac{(2n+1) \pi y}{w} +
\cosh \rho_n y  \ \cos  \frac{(2n+1) \pi x}{w} \right]
\eeq
where $\rho_n \equiv \sqrt{\frac{(2n+1)^2 \pi^2}{w^2} - \frac{2m E_0}{\hbar^2}}$. Notice that there is no need of writing the remaining solutions, 
since they are obtained from $\psi_I(x,y)$ by means of rotations and reflections. The unknown coefficients $b_n$ are obtained by imposing 
the continuity of the normal derivative of $\psi(x,y)$ at the border between the regions I and V.Using $N=8$ we have obtained $E=0.849555$
(using as before $w=2$ and $\hbar=m=1$), corresponding to $k w = \frac{w \sqrt{2 m E_0}}{\hbar}\approx 0.829833 \pi$, which is about $2 \%$ larger
than the result previously obtained (this is consistent with the result found in \cite{Avishai91}).

As we have seen previously the inclusion of a circular piercing is obtained by multiplying the original solution by a factor $\log r/r_0$, 
$r_0$ being the radius of the piercing. If this is correct, one should see that, as $r_0$ is sent to zero, the energy approaches the value in
the absence of piercing.

\begin{figure}
~\bigskip\bigskip
\begin{center}
\includegraphics[width=8cm]{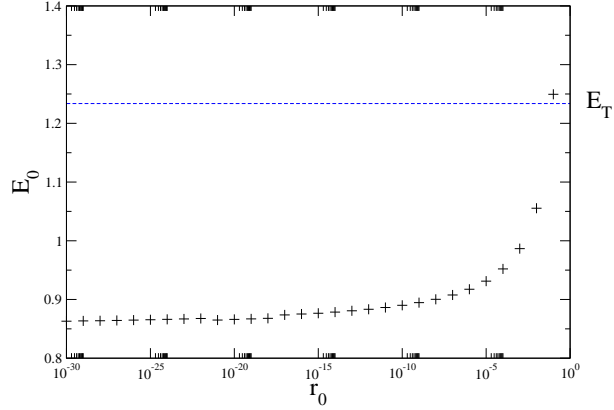}
\caption{Energy of the ground state of a crossed wire with a piercing in the origin as a function of the  radius $r_0$. The horizontal line
is the threshold of the continuum. (color online)}
\bigskip
\label{fig7}
\end{center}
\end{figure}

\begin{figure}
~\bigskip\bigskip
\begin{center}
\includegraphics[width=8cm]{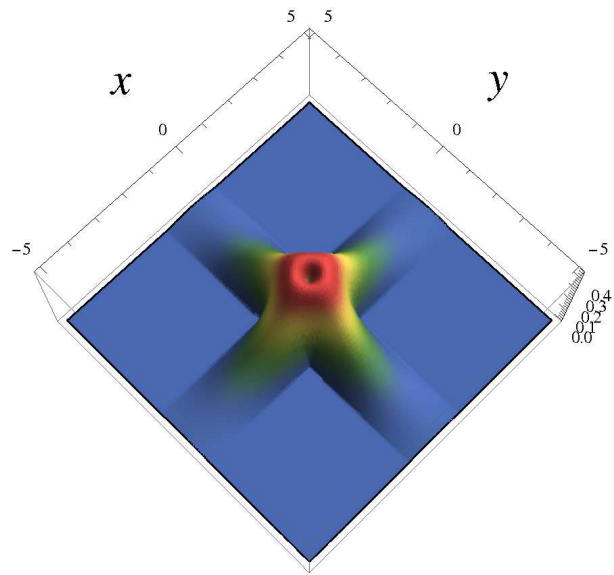}
\caption{Asymptotic wave function of the ground state of a crossed wire with a piercing of radius $r_0 = 10^{-3}$. (color online)}
\bigskip
\label{fig8}
\end{center}
\end{figure}

In Fig.~\ref{fig7} we show the energy of the ground state of the crossed wire with a piercing of radius $r_0$ in the origin as a function of 
$r_0$ itself. Using once more a least square procedure we find that in the limit $r_0 \rightarrow 0$, $E_0 \approx 0.848$, which is remarkably 
close to the value without piercing. The critical value of the piercing radius for which the threshold energy is reached is $r_0^{crit} \approx 0.0899$.
In Fig.~\ref{fig8} we show the wave function of the ground state of the crossed wire with a piercing of radius $r0=10^{-3}$ at the origin, obtained
using the expansion in terms of the elementary solutions.

\section{Conclusions}
\label{conclu}

In this paper  we have proved that the spectrum of the $d$-dimensional Schr\"odinger equation does not change when piercings of infinitesimal
size are added to the $d$-dimensional domain. The example of the simple harmonic oscillator in $d$ dimensions is worked out and the expected 
results are obtained both analytically (for a single piercing in the origin) and numerically (up to $20$ piercings). Using these results we have 
considered a two dimensional parabolic quantum dot and we have calculated the energy of the ground state to first order in perturbation theory, 
up to the leading finite size correction in the piercing radius. We have found that the interaction energy of the quantum dot is lower for 
piercings of finite size, and that it can dominate the corresponding finite size correction to the single particle energy above a critical coupling. 
In our calculations the critical coupling turns out too be large to trust a first order perturbative result. This outcome should motivate 
a further study of this system, either involving higher order perturbative corrections or a variational calculation, which we hope to carry out
soon. Another application considered in the present paper is to configurations supporting bounds states in the continuum, such as wires with
crossings and bendings. We have explicitly shown that the inclusion of piercings to these systems does not alter the energy of the ground state,
contrary to naive expectations and in perfect accord with our general considerations. 

A further remark that we wish to make concerns the Casimir effect on domains with piercings: at least in the case of non interacting fields, and
for piercings of infinitesimal size, our results imply that no net effect should appear when piercings are added to a domain. In presence of
interactions, whose study is certainly a formidable task, the inclusion of piercings may affect the spectrum thus leading to a net effect.

\acknowledgments
The author ackowledges the support received by SEP, through Cuerpo Academico UCOL-CA56.

\end{document}